\documentclass[10pt,conference]{IEEEtran}
\IEEEoverridecommandlockouts
\usepackage{cite}
\usepackage{amsmath,amssymb,amsfonts}
\usepackage{algorithmic}
\usepackage{graphicx}
\usepackage{textcomp}
\usepackage{xcolor}
\usepackage{tabularx}
\usepackage{color}
\usepackage{booktabs}
\usepackage{multirow}
\usepackage{subfigure}
\usepackage{tikz}
\def\BibTeX{{\rm B\kern-.05em{\sc i\kern-.025em b}\kern-.08em
    T\kern-.1667em\lower.7ex\hbox{E}\kern-.125emX}}
\begin{document}

\title{\huge{Multi-task Single Channel Speech Enhancement Using Speech Presence Probability As A Secondary Task Training Target}
\thanks{This work is supported by the National Key Research Project of China under Grant No. 2017YFF0210903 and the National Natural Science Foundation of China under Grant No. 61371147.}
}

\author{\IEEEauthorblockN{Lei Wang}
\IEEEauthorblockA{\textit{Dept. of Electronic Engineering} \\
\textit{Shanghai Jiao Tong University}\\
Shanghai, China \\
wang\_lei@sjtu.edu.cn}
\and
\IEEEauthorblockN{Jie Zhu}
\IEEEauthorblockA{\textit{Dept. of Electronic Engineering} \\
\textit{Shanghai Jiao Tong University}\\
Shanghai, China \\
zhujie@sjtu.edu.cn}
\and
\IEEEauthorblockN{Ina Kodrasi}
\IEEEauthorblockA{\textit{Speech and Audio Processing Group} \\
\textit{Idiap Research Institute}\\
Martigny, Switzerland \\
ina.kodrasi@idiap.ch}
}

\maketitle

\begin{abstract}
  To cope with reverberation and noise in single channel acoustic scenarios, typical supervised deep neural network~(DNN)-based techniques learn a mapping from reverberant and noisy input features to a user-defined target.
  Commonly used targets are the desired signal magnitude, a time-frequency mask such as the Wiener gain, or the interference power spectral density and signal-to-interference ratio that can be used to compute a time-frequency mask.
  In this paper, we propose to incorporate multi-task learning in such DNN-based enhancement techniques by using speech presence probability (SPP) estimation as a secondary task assisting the target estimation in the main task.
  The advantage of multi-task learning lies in sharing domain-specific information between the two tasks (i.e., target and SPP estimation) and learning more generalizable and robust representations.
  To simultaneously learn both tasks, we propose to use the adaptive weighting method of losses derived from the homoscedastic uncertainty of tasks.
  Simulation results show that the dereverberation and noise reduction performance of a single-task DNN trained to directly estimate the Wiener gain is higher than the performance of single-task DNNs trained to estimate the desired signal magnitude, the interference power spectral density, or the signal-to-interference ratio.
  Incorporating the proposed multi-task learning scheme to jointly estimate the Wiener gain and the SPP increases the dereverberation and noise reduction further.
\end{abstract}

\label{intro}

\section{Introduction}
\label{intro}
In many speech communication applications, the recorded microphone signal is inevitably corrupted with late reverberation and noise, which can be detrimental to speech quality and intelligibility and to the accuracy of speech recognition systems~\cite{Beutelmann_JASA_2006 
, Warzybok_IWAENC_2014}.
The goal of single channel speech enhancement is to recover the desired signal while suppressing the interference, i.e., late reverberation and noise.
Single channel speech enhancement has been traditionally approached using spectral enhancement techniques~\cite{gerkmann2011unbiased, Cauchi_EURASIP_2015
} or probabilistic modeling-based techniques~\cite{Attias_Neural_2000, Yoshioka_ITASLP_2009}.
In recent years however, successful contributions based on data-driven approaches such as deep neural networks~(DNNs) have been proposed~\cite{Wang_ITASLP_2018}.

Typical supervised DNN-based techniques for single channel speech enhancement learn a mapping from reverberant and noisy input features to a user-defined target~\cite{Wang_ITASLP_2018}.
Depending on the definition of the target, such techniques can be broadly categorized into magnitude estimation~\cite{Han_ITASLP_2015, 
Wu_Asia_2016, 
Wu_ITASLP_2017} and mask estimation techniques~\cite{Han_JASA_2012, Wang_ITASLP_2014, Wang_ITASLP_2020}.
Magnitude estimation techniques aim at estimating the desired signal spectral magnitude.
The enhanced signal is then obtained by combining the estimated magnitude with the phase of the recorded microphone signal.
Mask estimation techniques on the other hand aim at estimating a time-frequency mask such as the Wiener gain.
The enhanced signal is then obtained by applying the estimated time-frequency mask to the recorded microphone signal.
Instead of directly estimating the time-frequency mask, indirect mask estimation techniques have been recently proposed in~\cite{kodrasi2018single, Nicolson_SC_2019}, where the interference power spectral density~(PSD) or the signal-to-interference ratio~(SIR) are estimated.
The estimated interference PSD or SIR can then used to define a time-frequency mask to recover the enhanced signal.

To improve the generalization performance of such DNN-based enhancement techniques, in this paper we propose to incorporate multi-task learning~\cite{Caruana1998Multitask}, which means using one network to estimate multiple targets simultaneously.
Multi-task learning has been successfully applied in various areas such as computer vision~\cite{Wang_CVPR_2009} or natural language processing~\cite{Collobert_ICML_2008}, in this paper it is incorporated for DNN-based single channel speech enhancement.
Multi-task learning improves learning efficiency and generalization performance by using shared representations to jointly learn multiple related tasks, such that what is learned from one task can help learning and generalization in another task.
To incorporate multi-task learning in supervised DNN-based single channel speech enhancement techniques, we propose to use speech presence probability (SPP) estimation as a secondary task.
SPP is a useful parameter in traditional single channel speech enhancement techniques for accurately estimating the interference PSD, and hence, for improving the speech enhancement performance~\cite{gerkmann2011unbiased,2010Simultaneous,2011Noise}.
Consequently, we expect that the incorporation of SPP estimation as a secondary task results in learning more robust representations for the primary target (i.e., desired signal magnitude, time-frequency mask, interference PSD, or SIR) estimation task.
To simultaneously learn both tasks, we propose to use the adaptive weighting method of losses derived from the homoscedastic uncertainty of tasks in \cite{kendall2018multi}.

 \section{DNN-based single channel enhancement}
\label{sec2}

We consider a reverberant and noisy microphone system with a single speech source and a single microphone. 
In the short-time Fourier transform~(STFT) domain, the received microphone signal $Y(k,l)$ at frequency bin $k$ and time frame index $l$ can be written as
\begin{equation}
\label{2}
Y(k,l) = X(k,l)+ \underbrace{R(k,l)+N(k,l)}_{I(k,l)},\\
\end{equation}
with $X(k,l)$ being the direct and early reverberation component, $R(k,l)$ being the late reverberation component, $N(k,l)$ being the additive noise component, and $I(k,l)$ denoting the total interference component (i.e., late reverberation and noise).
Assuming that $X(k,l)$ and $I(k,l)$ are uncorrelated, the PSD of the microphone signal $Y(k,l)$ is given by
\begin{equation}
  \Phi^2_y(k,l) = {\cal{E}}\{|Y(k,l)|^2 \} = \Phi^2_x(k,l) + \Phi^2_i(k,l),
\end{equation}
with ${\cal{E}}$ denoting the expected value operator and $\Phi^2_x(k,l)$ and $\Phi^2_i(k,l)$ denoting the PSDs of $X(k,l)$ and $I(k,l)$, respectively.

Since early reverberation is desirable~\cite{Bradely_JASA_2003}, the objective of speech enhancement is to recover an estimate of the direct and early reverberation component $X(k,l)$.
Typical DNN-based techniques aiming to recover $X(k,l)$ are trained to learn a mapping from reverberant and noisy input features to a user-defined target.
Depending on the target definition, such techniques can be broadly categorized into magnitude estimation~\cite{Han_ITASLP_2015, Wu_Asia_2016, Wu_ITASLP_2017} and mask estimation techniques~\cite{Han_JASA_2012, Wang_ITASLP_2014, Wang_ITASLP_2020, kodrasi2018single, Nicolson_SC_2019}.
Mask estimation techniques can be additionally categorized into 3 subcategories, i.e., directly time-frequency mask estimation~\cite{Han_ITASLP_2015, Wu_Asia_2016, Wu_ITASLP_2017}, interference PSD estimation required to compute a time-frequency mask~\cite{kodrasi2018single}, a priori SIR estimation required to compute a time-frequency mask~\cite{Nicolson_SC_2019}.
These techniques differ not only in terms of the target definition, but also in terms of the used input features and DNN architectures.
However, to provide a systematic review and compare the performance for different targets in Section~\ref{exp}, in this paper we consider only different target definitions for standard feed-forward DNN architectures with temporal context depicted in Figs.~\ref{n1}(a) and~\ref{n1}(b).
Next, a brief overview of the considered input and target definitions for such DNNs is provided.

\subsection{Magnitude estimation}
When estimating the desired signal magnitude, the DNN target vector can be defined as the $K$--dimensional vector constructed using the spectral magnitude of $X(k,l)$ at time frame $l$ across all frequency bins $K$, i.e.,
\begin{equation}
  \mathbf{ x}(l)=[|X(1,l)| \; |X(2,l)| \; \ldots \; |X(K,l)|]^T.
\end{equation}
To incorporate temporal context, the DNN input vector can be defined as the $K(2T+1)$--dimensional vector constructed by concatenating the spectral magnitude of $Y(k,l)$ from the past and future $T$ time frames across all frequency bins $K$, i.e.,
\begin{eqnarray}
\label{4}
  \mathbf{y}(l)  & =  [|Y(1,l-T)| \; \ldots \; |Y(K,l-T)| \ldots  \\ \nonumber
                & \ldots |Y(1,l+T)| \ldots |Y(K,l+T)|]^T.
\end{eqnarray}
Using the estimated spectral magnitude $|\hat{X}(k,l)|$, the enhanced signal can be obtained as $\hat{X}_{\text{mag}}(k,l)=\frac{|\hat{X}(k,l)|}{|Y(k,l)|}Y(k,l)$.


\subsection{Mask estimation}
Although different time-frequency masks have been investigated in the literature~\cite{Wang_ITASLP_2014,Wang_ITASLP_2018}, the commonly used Wiener gain is considered in this paper.
With the a priori SIR $\xi(k,l)$ defined as $\xi(k,l) = \frac{\Phi^2_x(k,l)}{\Phi^2_i(k,l)}$,
the Wiener gain can be computed as
\begin{equation}
  \label{6}
G(k,l)=\frac{\xi(k,l)}{\xi(k,l)+1}.
\end{equation}

\subsubsection{Direct mask estimation}
When directly estimating the Wiener gain, the DNN target vector can be defined as the $K$--dimensional vector constructed using the gain $G(k,l)$ at time frame $l$ across all frequency bins $K$, i.e.,
\begin{equation}
  \label{eq: wg}
  \mathbf{G}(l)=[G(1,l) \; G(2,l) \; \ldots \; G(K,l)]^T,
\end{equation}
whereas the DNN input vector can be defined as the $K(2T+1)$--dimensional vector $\mathbf{y}(l)$ in~(\ref{4}).
Using the estimated Wiener gain $\hat{G}(k,l)$, the enhanced signal can be obtained as $\hat X_{\text{gain}}(k,l)=\hat G(k,l)Y(k,l)$.

\subsubsection{Interference PSD estimation}
Instead of directly estimating the gain in~(\ref{6}), in~\cite{kodrasi2018single} it has been proposed to use a DNN for estimating the interference PSD $\Phi^2_i(k,l)$. 
Hence, the DNN target vector can be defined as the $K$--dimensional vector constructed using the interference PSD $\Phi_i^2(k,l)$ at time frame $l$ across all frequency bins $K$, i.e.,
\begin{equation}
  \label{eq: phii}
  \mathbf{ \Phi}^2_i(l)=[\Phi^2_i(1,l) \; \Phi^2_i(2,l) \; \ldots \; \Phi^2_i(K,l)]^T.
\end{equation}
Further, the DNN input vector can be defined as the $K(2T+1)$--dimensional vector constructed by concatenating the microphone signal PSD $\Phi^2_y(l)$ from the past and future $T$ time frames as in~(\ref{4}), i.e.,
\begin{eqnarray}
\label{11}
\mathbf{\Phi}^2_y(l)= & [ |\Phi^2_y(1,l-T)| \; \ldots \; |\Phi^2_y(K,l-T)| \ldots \\ \nonumber
& \ldots |\Phi^2_y(1,l+T)| \; \ldots \; |\Phi^2_y(K,l+T)|]^T.
\end{eqnarray}
To compute the enhanced signal, first the estimated interference PSD $\hat{\Phi}^2_i(k,l)$ is used to obtain an estimate of the a priori SIR $\hat{\xi}_{\text{psd}}(k,l)$ based on the decision directed approach~\cite{ephraim1984speech}.
The estimated a priori SIR $\hat{\xi}_{\text{psd}}(k,l)$ is then exploited to compute the Wiener gain $\hat{G}_{\text{psd}}$ as in~(\ref{6}), yielding the enhanced signal $\hat X_{\text{psd}}(k,l)=\hat{G}_{\text{{psd}}}Y(k,l)$.

\subsubsection{SIR estimation}
Instead of directly estimating the Wiener gain in~(\ref{6}), in~\cite{Nicolson_SC_2019} it has been proposed to use a DNN for estimating the SIR $\xi(k,l)$.
Hence, the DNN target vector can be constructed as
\begin{equation}
  \label{eq: isir}
  \boldsymbol{\xi}(l)=[\xi(1,l) \; \xi(2,l) \; \ldots \; \xi(K,l)]^T,
\end{equation}
whereas the DNN input vector is the $K(2T+1)$--dimensional vector $\mathbf y(l)$ defined in~(\ref{4}).
To compute the enhanced signal, the estimated a priori SIR is used to compute the Wiener gain $\hat{G}_{\text{sir}}$ as in~(\ref{6}), yielding $\hat X_{\text{sir}}(k,l)=\hat{G}_{\text{{sir}}}Y(k,l)$.

\begin{figure*}[htbp] 
\centering
\subfigure{\label{h2}
\begin{minipage}[c]{0.12\textwidth}
\centering
\includegraphics[width=1\textwidth]{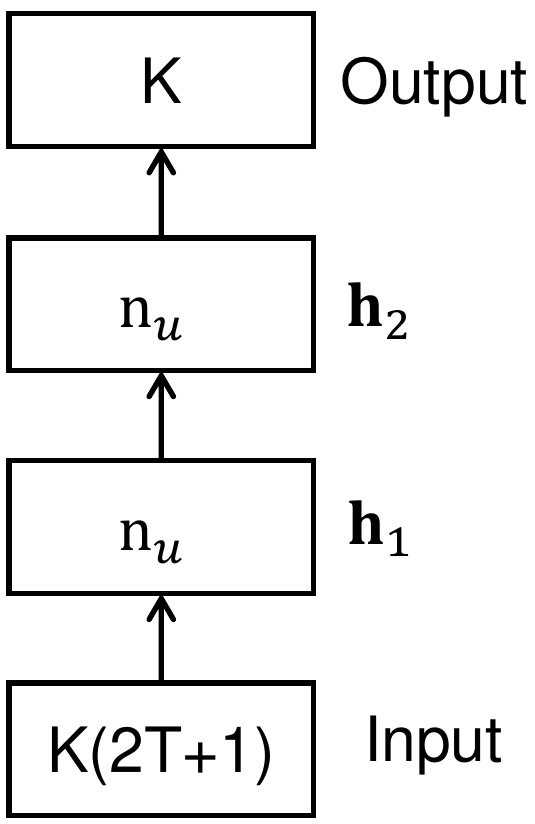}
\end{minipage}
}
\subfigure{\label{h3}
\begin{minipage}[c]{0.12\textwidth}
\centering
\includegraphics[width=1\textwidth]{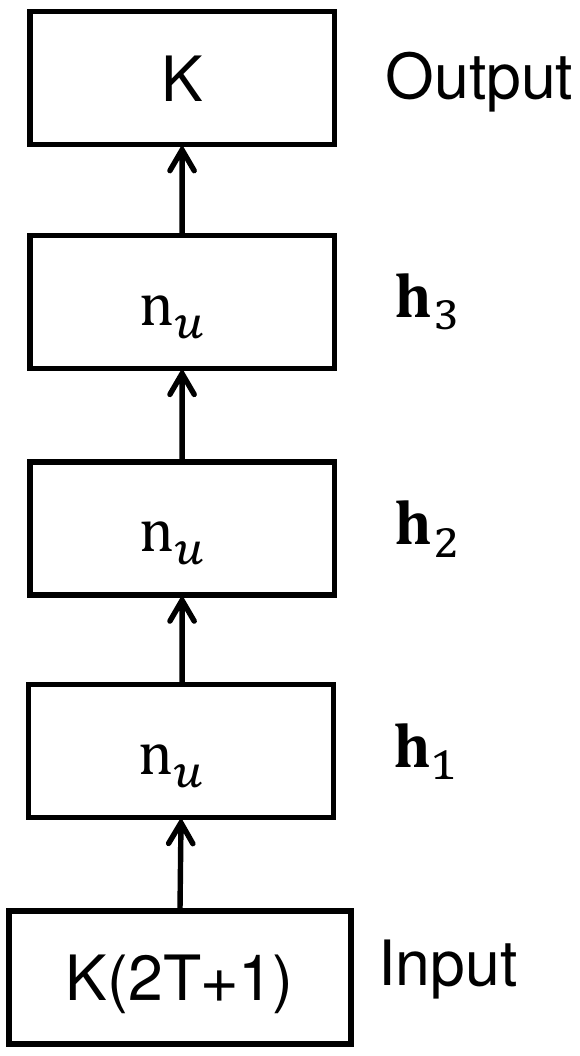}
\end{minipage}
}
\subfigure{\label{mh2}
\begin{minipage}[c]{0.22\textwidth}
\centering
\includegraphics[width=1\textwidth]{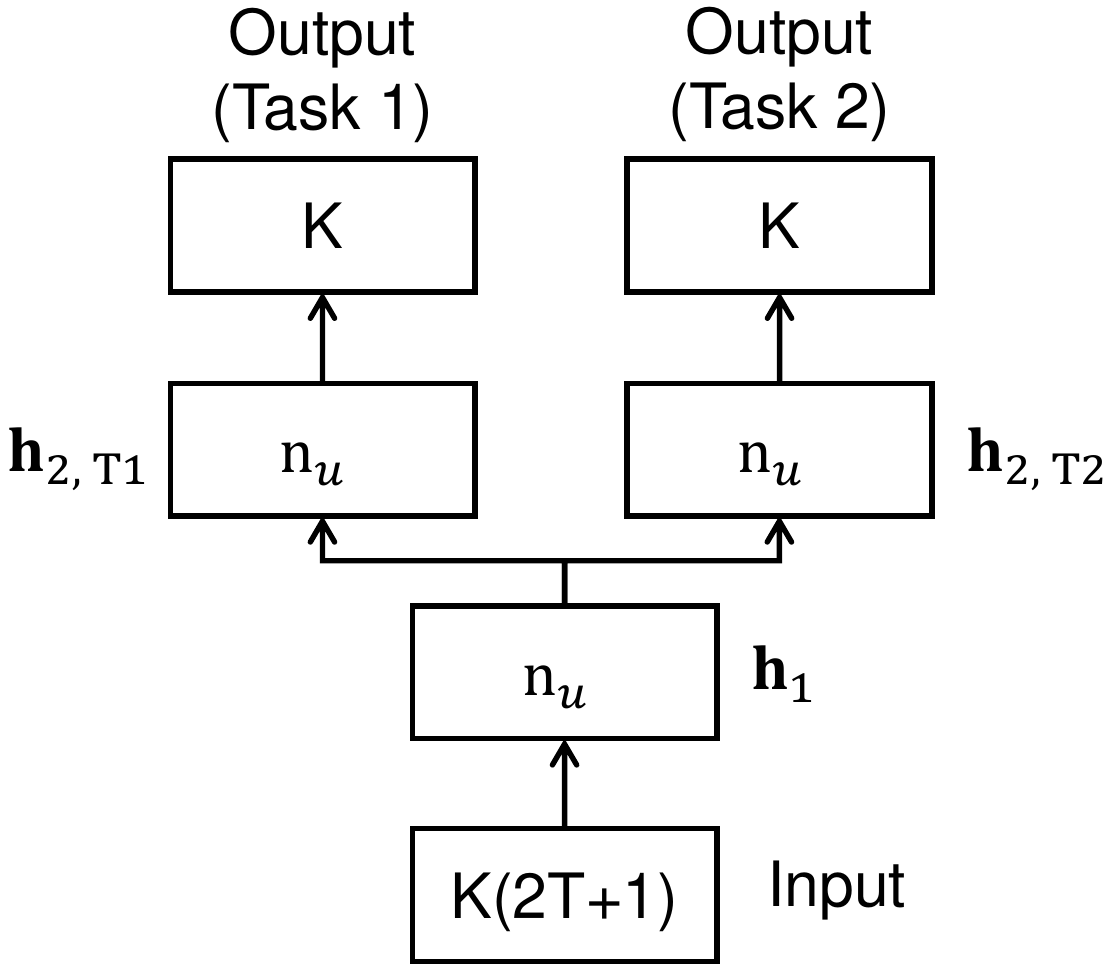}
\end{minipage}
}
\subfigure{\label{mh32}
\begin{minipage}[c]{0.22\textwidth}
\centering
\includegraphics[width=1\textwidth]{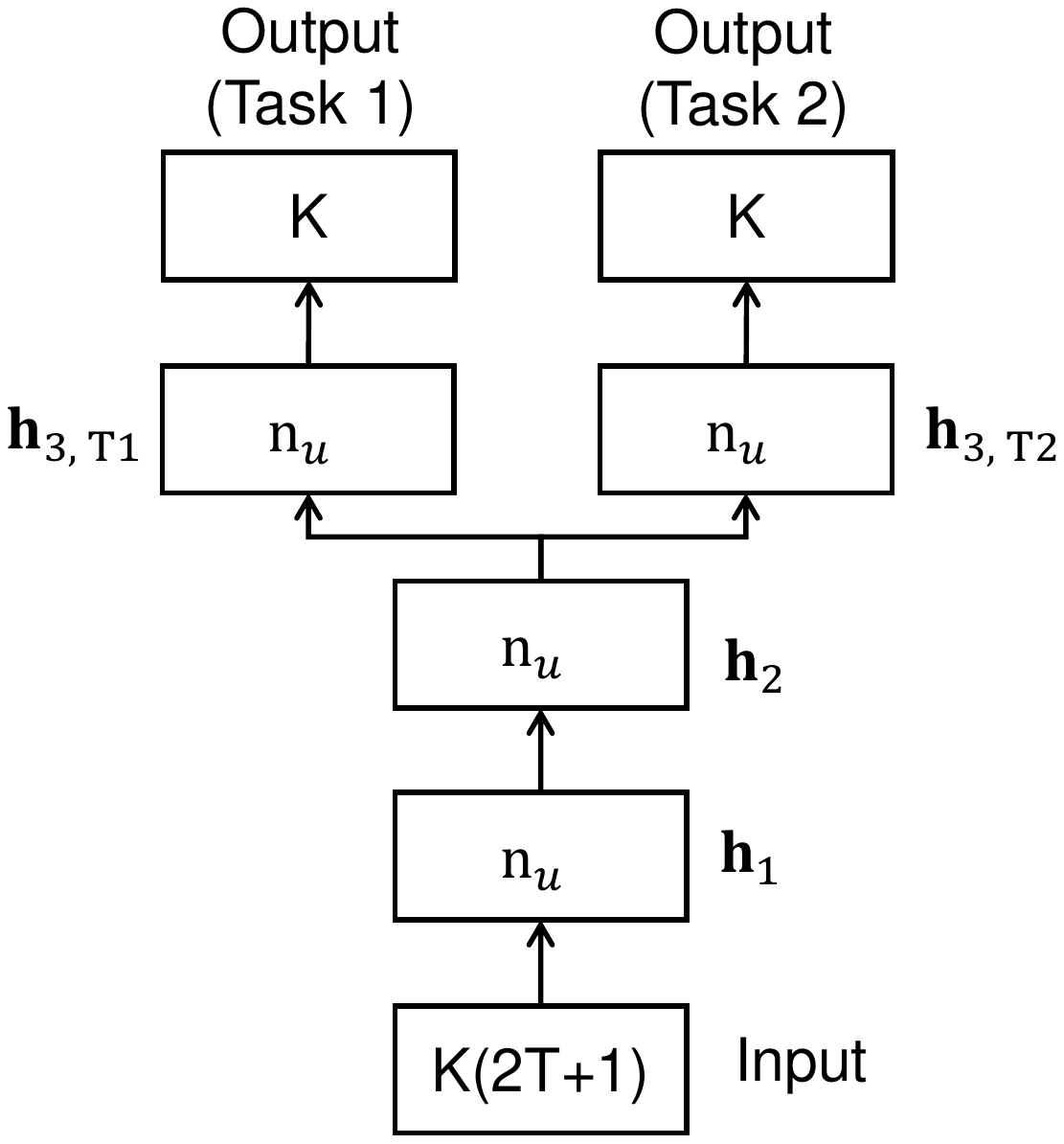}
\end{minipage}
}
\subfigure{\label{mh31}
\begin{minipage}[c]{0.22\textwidth}
\centering
\includegraphics[width=1\textwidth]{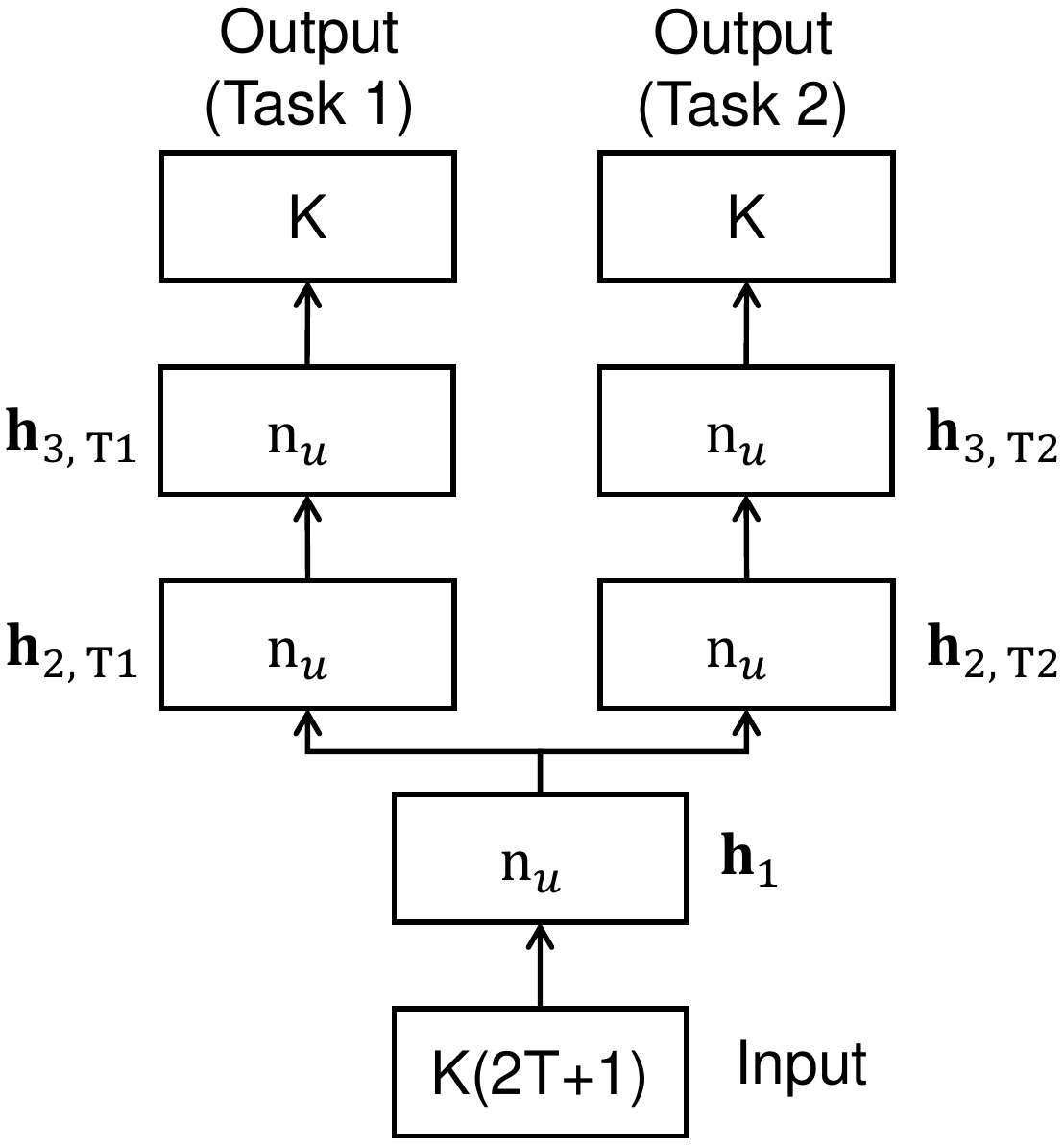}
\end{minipage}
}
\centering
\centerline{(a) \qquad\qquad\qquad(b) \qquad\qquad\qquad\qquad\qquad\quad (c) \qquad\qquad\qquad\qquad\qquad\quad (d) \qquad\qquad\qquad\qquad\qquad\qquad\ (e)\qquad\quad\ }
\caption{Schematic illustration of the considered DNN architectures: (a) single-task estimation with two layers, (b) single-task estimation with three layers, (c) multi-task estimation with one shared layer followed by one task-specific layer, (d) multi-task estimation with two shared layers followed by one task-specific layer, and (e) multi-task estimation with one shared layer followed by two task-specific layers.}
\label{n1}
\end{figure*}

\section{Multi-task learning for \\ DNN-based speech enhancement}
\label{sec: mtl}

In this section, we propose to increase the generalization performance of the DNN-based speech enhancement techniques reviewed in Section~\ref{sec2} by incorporating multi-task learning, where multiple targets are learned simultaneously in one network.
Instead of using a single-task DNN that only estimates the user-defined target (i.e., desired signal magnitude, time-frequency mask, interference PSD, or SIR), we propose to use a multi-task DNN that additionally estimates the SPP.
The SPP is a useful parameter in single channel speech enhancement for accurately tracking the interference PSD, and hence, for improving the speech enhancement performance~\cite{gerkmann2011unbiased}.
We hypothesize that jointly learning to estimate the user-defined target and the SPP through shared DNN layers within a multi-task learning framework yields more robust and generalizable representations for the primary task~(i.e., estimating the user-defined target).
Assuming that the desired signal and interference STFT coefficients are complex Gaussian distributed, the SPP can be computed as~\cite{gerkmann2011unbiased}
\begin{equation}
\label{17}
\text{SPP}(k,l)=\left(1+\frac{P(\mathcal{H}_0)}{P(\mathcal{H}_1)}(1+\xi_{\mathcal{H}_1})\rm e^{-\frac{|Y(k,l)|^2}{\Phi_i^2(k,l)}\frac{\xi_{\mathcal{H}_1}}{1+\xi_{\mathcal{H}_1}}}\right)^{-1}\!\!\!\!\!\!,
\end{equation}
where $P(\mathcal{H}_1)$ and $P(\mathcal{H}_0)$ are the prior probabilities of speech presence and absence and $\xi_{\mathcal{H}_1}$ denotes the typical a priori SIR when speech is present.
In line with the target definitions in Section~\ref{sec2}, the target vector for SPP estimation is given by
\begin{equation}
  \label{eq: spp}
 { \textbf{SPP}}(l)=[\text{SPP}(1,l) \; \text{SPP}(2,l) \; \ldots \; \text{SPP}(K,l)]^T.
\end{equation}

\begin{table*}[t!]
  \caption{Performance of single-task estimation of different targets on the test reverberant, noisy, and unseen noisy datasets.}
  \label{tbl: st}
  \centering
  \begin{tabularx}{\linewidth}{X|rrrr|rrrr|rrrr}
    \toprule
    \def\tabcolsep{3.5pt}
    & \multicolumn{4}{c|}{Reverberant} &  \multicolumn{4}{c|}{Noisy} & \multicolumn{4}{c}{Unseen Noisy}\\
    Measure & $\hat{X}_{\text{mag}}$ & $\hat{X}_{\text{gain}}$ & $\hat{X}_{\text{psd}}$ & $\hat{X}_{\text{sir}}$ & $\hat{X}_{\text{mag}}$ & $\hat{X}_{\text{gain}}$ & $\hat{X}_{\text{psd}}$ & $\hat{X}_{\text{sir}}$ & $\hat{X}_{\text{mag}}$ & $\hat{X}_{\text{gain}}$ & $\hat{X}_{\text{psd}}$ & $\hat{X}_{\text{sir}}$ \\
    \toprule
    $\Delta$PESQ & 0.14 & \bf 0.23 & 0.13 & 0.14  &0.04& \bf 0.22&0.13&0.10 & 0.00& \bf 0.23&0.16&0.12 \\
    \hline
    $\Delta$fwSSNR & 1.80 & \bf 2.27 & 1.07 & 1.31 & 1.77 & \bf 2.68 & 1.64 & 1.19 & 1.06 & \bf 2.38 & 1.71 & 0.98 \\
    \bottomrule
  \end{tabularx}
\end{table*}

Figs.~\ref{n1}(c)--\ref{n1}(e) depict examples of the considered DNN architectures for jointly learning two different tasks, with the first task being the estimation of a target vector as presented in Section~\ref{sec2} and the second task being the estimation of the SPP in~(\ref{eq: spp}).
In Fig.~\ref{n1}(c) both tasks share one hidden layer followed by a task-specific layer, in Fig.~\ref{n1}(d) both tasks share two hidden layers followed by a task-specific layer, whereas in Fig.~\ref{n1}(e) both tasks share one hidden layer followed by two task-specific layers.
To train these architectures, the loss function can be defined as a weighted sum of the task-specific loss functions~\cite{kendall2018multi}, i.e.,
\begin{equation}
\label{18}
\mathcal{L}_{\text{fixed}}(\mathbf{W})=\lambda_1 \mathcal{L}_1(\mathbf{W}) +\lambda_2 \mathcal{L}_2(\mathbf{W}),
\end{equation}
with $\mathcal{L}_1$ being the loss function for estimating a target vector from Section~\ref{sec2}, $\mathcal{L}_2$ being the loss function for estimating the SPP in~(\ref{eq: spp}), $\lambda_1$ and $\lambda_2$ being user-defined weighting scalars, and $\textbf{W}$ being the model parameters.
When using the loss function in~(\ref{18}), the performance of the model can be sensitive to the values of $\lambda_1$ and $\lambda_2$ and finding optimal values can be expensive~\cite{kendall2018multi}.
To avoid tuning $\lambda_1$ and $\lambda_2$, we propose to use the adaptive loss function derived in~\cite{kendall2018multi} to automatically weight the task-specific loss functions, i.e.,
\begin{equation}
\label{19}
\mathcal{L}_{\text{ada}}(\textbf{W},\sigma_1,\sigma_2)=\frac{1}{\sigma_1^2}\mathcal{L}_1(\textbf{W})+\frac{1}{\sigma_2^2}\mathcal{L}_2(\textbf{W})+\rm{log}\,\sigma_1\sigma_2,
\end{equation}
where $\sigma_1$ and $\sigma_2$ are scalars jointly learned with the model parameters $\mathbf{W}$.
Although not presented in this paper due to space constraints, using~(\ref{19}) yields a better performance than using~(\ref{18}) for several user-defined $\lambda_1$ and $\lambda_2$ for the reverberant and noisy acoustic scenarios considered in Section~\ref{exp}.

\section{Simulation results}
\label{exp}
In this section, the performance of all single-task techniques discussed in Section~\ref{sec2} is first compared on the same datasets and DNN architectures.\footnote{To the best of our knowledge, only the performance of magnitude and direct mask estimation techniques has been compared on the same datasets and DNN architectures in~\cite{Wang_ITASLP_2014}, while the performance of the more recently proposed interference PSD and SIR estimation techniques has not been considered.}
Further, the performance of the proposed multi-task framework for joint direct mask and SPP estimation is investigated.

\subsection{Datasets}
Two datasets are considered, i.e., a reverberant dataset where the interference consists of different reverberation levels and a reverberant and noisy dataset (referred to as a noisy dataset) where the interference consists of a fixed reverberation level and varying levels and types of noise.
As clean speech material, we have used the TIMIT database~\cite{timit}.

To generate the reverberant dataset, clean speech files are convolved with measured room impulse responses~(RIRs) with reverberation times ranging from $200$~ms to $1$~s.
For the reverberant training, validation, and test sets we have used $500$, $200$, and $200$ clean speech files and $16$, $8$, and $8$ RIRs, respectively, with no overlap between files for different sets.

To generate the noisy dataset, clean speech files are firstly convolved with one measured RIR and corrupted with different noise types from the DEMAND database~\cite{demand}.
For the training, validation, and test sets we have used $250$, $100$, and $100$ clean speech files convolved with an RIR with reverberation time $580$ ms, $570$ ms, and $560$ ms, respectively.
As before, there is no overlap between the clean speech files and the RIRs for different sets.
Further, for the training, validation, and test sets, $5$ different noise types at $3$ different broadband signal-to-noise ratio~(SNR) are added to the reverberant signals, with SNR $\in \{ -5 \text{dB}, 0 \text{dB}, 5 \text{dB} \}$.
To analyze the generalization capabilities of the proposed models, an unseen noisy test set is also generated by adding $3$ unseen noise types at unseen broadband SNRs to the test reverberant signals, with SNR$\in\!\{ -3 \text{dB}, 3 \text{dB}, 10 \text{dB} \}$.

\subsection{Parameters, network architectures, and measures}
{\emph{Parameters.} \enspace
Signals are processed in the STFT domain using a weighted overlap-add framework with a tight analysis window of $256$ samples and an overlap of $50$\%.
Considering only half of the spectrum, the number of frequency bins is $K = 129$.
Further, the number of time frames used for temporal context is $T = 3$.
To compute the PSDs required in~(\ref{6})--(\ref{17}), we use recursive averaging with a smoothing factor of $0.85$.
To compute the SPP in~(\ref{17}) we use $P(\mathcal{H}_1) = 0.5$, $P(\mathcal{H}_0) = 0.5$, and $10 \log_{10}\xi_{\mathcal{H}_1} = 15$~dB.

{\emph{Network architectures.} \enspace
  As previously mentioned, the network architectures considered for the single- and multi-task techniques are depicted in Fig.~\ref{n1}.
  For all architectures, we use rectifying linear unit (ReLU) as non-linearity on all hidden layers. 
  For estimating an unbounded target (i.e., the desired signal magnitude, the interference PSD, or the SIR), there is no non-linearity on the output layer.  
  For estimating the Wiener gain or the SPP which are bounded between $0$ and $1$, a sigmoid non-linearity is used on the output layer.
  Mean square error is used as the loss function for training the single-task architectures in Figs.~\ref{n1}(a), (b) and as the loss function ${\cal{L}}_1$ for training the multi-task architectures in Figs.~\ref{n1}(c)--\ref{n1}(e).
  Cross-entropy loss is used as the loss function ${\cal{L}}_2$ for training the multi-task architectures in Figs.~\ref{n1}(c)--\ref{n1}(e).
  All considered architectures are trained for different number of hidden units $n_u \in \{$500, 1000, 1500$\}$ using the Adam optimizer with different hyper-parameters, i.e., learning rate $l_r \in \{0.001, 0.0001 \}$ and weight decay $w_d \in \{0, 0.001 \}$.
  After training for $200$ epochs, the model parameters corresponding to the epoch with the lowest validation error (out of all considered architectures, $n_u$, $l_r$, and $w_d$) are used as the final model parameters.

  {\emph{Measures.} \enspace
    The dereverberation and denoising performance is measured by the improvement in perceptual evaluation of speech quality ($\Delta$PESQ)~\cite{PESQ} and frequency-weighted segmental signal to noise ratio ($\Delta$fwSSNR)~\cite{Hu_ITASLP_2008} between the processed and recorded microphone signals.

\subsection{Single-task performance}
\label{stl}
The performance of the techniques in Section~\ref{sec2} is compared on the test reverberant, noisy, and unseen noisy datasets.
As previously mentioned, for each technique, the two- and three-layer networks in Figs.~\ref{n1}(a) and~\ref{n1}(b) are trained for all considered hyper-parameters and the final network is selected as the one yielding the minimum validation loss.

Table~\ref{tbl: st} presents the performance on all considered test datasets, with the presented performance measures averaged over all utterances in the respective datasets.
It can be observed that the considered techniques generally yield an improvement in PESQ and fwSSNR on all datasets, with the direct mask estimation technique (i.e., $\hat{X}_{\text{gain}}$) yielding the best performance.
The advantageous performance of the direct mask estimation technique in comparison to magnitude estimation was already established in~\cite{Wang_ITASLP_2014}.
However, also the more recently proposed interference PSD and SIR estimation techniques show a lower dereverberation and noise reduction performance than the direct mask estimation technique on all datasets.

\subsection{Multi-task performance}
The results presented in Section~\ref{stl} confirm the advantageous performance of the direct mask estimation technique in comparison to other state-of-the-art techniques.
Hence, in the following, this technique is jointly used with SPP estimation within the proposed multi-task learning scheme.
As previously mentioned, the two- and three-layer networks depicted in Figs.~\ref{n1}(c)--~\ref{n1}(e) are trained for all considered hyper-parameters and the final network is selected as the one yielding the minimum validation loss.
The PESQ and fwSSNR improvement obtained on all considered datasets are shown in Table~\ref{tbl: fin}.
When comparing the presented $\Delta$PESQ and $\Delta$fwSSNR to the values in Table~\ref{tbl: st} (for $\hat{X}_{\text{gain}}(k,l)$), it can be seen that the proposed multi-task scheme improves the performance over single-task training on all datasets.
While a small difference can be observed in $\Delta$PESQ, a larger difference is observed in the presented $\Delta$fwSSNR values.

In summary, the presented results confirm that using a multi-task learning framework with SPP estimation improves the dereverberation and noise reduction performance of conventional single-task DNN-based enhancement techniques.
In the future, we will investigate the potential of incorporating parameters other than the SPP within the proposed multi-task learning framework.

\begin{table}[t!]
  \small
  \caption{Performance of the proposed multi-task framework for jointly estimating the Wiener gain ($\hat{X}_{\text{gain}}$) and the SPP on the reverberant, noisy, and unseen noisy test datasets.}
  \label{tbl: fin}
  \centering
  \def\tabcolsep{4pt}
  \begin{tabularx}{\linewidth}{Xrrr}
    \toprule
    & Reverberant & Noisy & Unseen Noisy \\
    \toprule
    $\Delta$PESQ&0.24&0.24&0.25\\
    \hline
    $\Delta$fwSSNR&2.40&3.11&2.74\\
    \bottomrule
  \end{tabularx}
  \vspace{-0.25cm}
\end{table}

\section{Conclusion}

In this paper, multi-task learning has been proposed to improve the performance of supervised DNN-based single channel speech enhancement techniques.
Instead of only estimating a user-defined target (e.g., the desired signal magnitude, a time-frequency mask such as the Wiener gain, the interference PSD, or the SIR), it has been proposed to also jointly estimate the SPP through shared DNN layers.
To simultaneously learn both tasks, we have used a recently proposed adaptive weighting method of losses derived from the homoscedastic uncertainty of tasks.
Simulation results on reverberant and noisy datasets show that jointly estimating the Wiener gain and the SPP within the proposed multi-task learning framework outperforms other state-of-the-art techniques.

\footnotesize
\bibliographystyle{IEEEbib}
\bibliography{refs}

\end{document}